%% file: main.tex
\documentclass[conference]{IEEEtran}
\IEEEoverridecommandlockouts
\usepackage{cite}
\usepackage{amsmath,amssymb,amsfonts}
\usepackage{algorithmic}
\usepackage{graphicx}
\usepackage{textcomp}
\usepackage{xcolor}
\def\BibTeX{{\rm B\kern-.05em{\sc i\kern-.025em b}\kern-.08em
    T\kern-.1667em\lower.7ex\hbox{E}\kern-.125emX}}

\usepackage[]{hyperref}

\usepackage[linesnumbered,ruled,vlined]{algorithm2e}
\usepackage{multirow}
\usepackage{pifont}
\usepackage{wrapfig}
\usepackage{booktabs}

\input{extra_command}

\begin{document}

\title{MetaDSE: A Few-shot Meta-learning Framework for Cross-workload CPU Design Space Exploration}

% \author{\IEEEauthorblockN{Runzhen XUE}

\author{Runzhen Xue$^{1, 2, *}$, Hao Wu$^{3, *}$, Mingyu Yan$^{1, 2, \dag}$,
Ziheng Xiao$^{1}$, Xiaochun Ye$^{1, 2}$, Dongrui Fan$^{1, 2}$ \\
\IEEEauthorblockA{
$^{1}$State Key Lab of Processors, Institute of Computing Technology, Chinese Academy of Sciences, Beijing, China; \\
$^{2}$University of Chinese Academy of Sciences, Beijing, China; \\
$^{3}$University of Electronic Science and Technology of China, Chengdu, China; \\
\{xuerunzhen21s, yanmingyu, xiaoziheng, yexiaochun, fandr\}@ict.ac.cn, 
wh.pyjnqd@gmail.com
}
}

\renewcommand{\thefootnote}{\fnsymbol{footnote}}

\maketitle

\footnotetext[1]{Runzhen Xue and Hao Wu contributed equally to this work.}
\footnotetext[2]{The corresponding author is Mungyu Yan (\textit{yanmingyu@ict.ac.cn}).}

\input{Tex/0-Abstract}

\input{Tex/1-Introduction}

\input{Tex/2-Background}
\input{Tex/3-Motivation}

\input{Tex/4-Methodology}
\input{Tex/5-Experiment_Setup}
\input{Tex/6-Experiment}

\input{Tex/7-Conclusion}

\section*{Acknowledgment}
This work was supported by National Key Research and Development Program (Grant No. 2022YFB4501400), the National Natural Science Foundation of China (Grant No. 62202451 and No. 62302477), CAS Project for Young Scientists in Basic Research (Grant No. YSBR-029), and CAS Project for Youth Innovation Promotion Association.

% The preferred spelling of the word ``acknowledgment'' in America is without 
% an ``e'' after the ``g''. Avoid the stilted expression ``one of us (R. B. 
% G.) thanks $\ldots$''. Instead, try ``R. B. G. thanks$\ldots$''. Put sponsor 
% acknowledgments in the unnumbered footnote on the first page.

\bibliographystyle{ieeetr}
\bibliography{ref}

\end{document}

%% file: extra_command.tex
\usepackage{xcolor}
\usepackage{soul}
\usepackage{comment}
\usepackage{amssymb}

\newcommand{\squishlist}{
   \begin{list}{$\bullet$}
    { \setlength{\itemsep}{0pt}      \setlength{\parsep}{0pt}
      \setlength{\topsep}{3pt}       \setlength{\partopsep}{0pt}
      \setlength{\listparindent}{-2pt}
      \setlength{\itemindent}{-5pt}
      \setlength{\leftmargin}{1em} \setlength{\labelwidth}{0em}
      \setlength{\labelsep}{0.5em} } }

\newcommand{\squishend}{
    \end{list}  }

\setuldepth{cat}

\newcommand{\note}[1]{{\color{magenta}$\square$}}
\newcommand{\acc}[2]{{\small{#1}\footnotesize{$\pm$#2}}}

%% file: Tex/0-Abstract.tex
\begin{abstract}
Cross-workload design space exploration (DSE) is crucial in CPU architecture design. Existing DSE methods typically employ the transfer learning technique to leverage knowledge from source workloads, aiming to minimize the requirement of target workload simulation. However, these methods struggle with overfitting, data ambiguity, and workload dissimilarity. 

To address these challenges, we reframe the cross-workload CPU DSE task as a few-shot meta-learning problem and further introduce MetaDSE. 
By leveraging model agnostic meta-learning, MetaDSE swiftly adapts to new target workloads, greatly enhancing the efficiency of cross-workload CPU DSE. Additionally, MetaDSE introduces a novel knowledge transfer method called the workload-adaptive architectural mask algorithm, which uncovers the inherent properties of the architecture.
Experiments on SPEC CPU 2017 demonstrate that MetaDSE significantly reduces prediction error by 44.3\% compared to the state-of-the-art.
MetaDSE is open-sourced and available at this \href{https://anonymous.4open.science/r/Meta_DSE-02F8}{anonymous GitHub.}

\end{abstract}

\begin{IEEEkeywords}
Design Space Exploration, Cross-workload, Prediction Model, Meta-learning
\end{IEEEkeywords}

% we introduce MetaDSE, a framework features a Workload-Adaptive Architecture Mapping (WAM) mechanism, enabling more accurate modeling of various workloads. By employing Model-Agnostic Meta-Learning (MAML), MetaDSE facilitates rapid adaptation to different workloads, enhancing both the efficiency and accuracy of CPU design space exploration (DSE) tasks.

%% file: Tex/1-Introduction.tex
\section{Introduction}
Design Space Exploration (DSE) is pivotal in modern CPU architecture design, enabling the optimization of configurations to balance multiple objectives like performance, power, and area (PPA)~\cite{PPA1, PPA2}. The increasing complexity of CPU architectures has led to an exponential expansion of the design space, presenting significant challenges in efficiently identifying optimal configurations.

Prior DSE frameworks~\cite{ActBoost, wangduo1, relative_work1, relative_work2, AttentionDSE, fan2024explainable, wangduo2, BoomExplorer1, BoomExplorer2} have utilized learning-based surrogate models to expedite the DSE process, achieving substantial time savings compared to exhaustive simulations. However, these workload-specific DSE frameworks rely heavily on extensive simulations to generate sufficient data for training machine learning models, necessitating the evaluation of numerous design points. This approach can be exceedingly time-intensive, especially for complex architectural spaces.

To mitigate this challenge, various studies~\cite{TrEnDSE, TrDSE, TrEE, PACT07, Microprocessor, ISCA19, TC10} have adopted the transfer learning technique to reduce simulation overhead. By leveraging previously collected data from source workloads, transfer learning enables the model to adapt to unseen target workloads, a process known as cross-workload DSE. 
Cross-workload DSE generally follows a two-stage training process: first, pre-training on data from source workloads, and then fine-tuning on limited data from the target workload.
However, current cross-workload DSE frameworks have notable limitations in both upstream pre-training stage and downstream adaptation stage:
\begin{itemize}
    \item \textbf{Upstream Pre-training Issues:} 
    Existing methods typically rely on machine learning based surrogate models~\cite{TrEnDSE, TrDSE, TrEE, ISCA19, TC10} combined with standard supervised learning, which are prone to overfitting the pre-training data. Additionally, the inherent data ambiguity across multiple workloads introduces optimization challenges for transfer learning, often hindering effective convergence.
    
    \item \textbf{Downstream Adaptation Issues:} 
    Given the high cost of simulations, training samples for new workloads are often limited, leading most existing frameworks to rely on knowledge transfer for adaptation. However, these approaches typically assume workload similarity as a prerequisite~\cite{TrEnDSE, Microprocessor, PACT07}, which is not always the case. When this assumption fails, the adaptation performance degrades significantly, limiting the effectiveness of traditional transfer learning methods in cross-workload CPU DSE tasks.
\end{itemize}

To tackle these issues, we rethink cross-workload CPU DSE as a few-shot problem and propose MetaDSE, an open-source framework featuring two key stages: upstream pre-training and downstream adaptation.
\textbf{For the upstream pre-training stage,} 
we adopt a meta-learning approach~\cite{MAML, meta_for_few_shot1, meta_for_few_shot2}, well-suited for addressing few-shot problems. 
Unlike normal supervised learning, meta-learning emphasizes "learning how to learn" across tasks, making it ideal for tackling data ambiguity in DSE. 
By treating each workload as a distinct task during meta-training, the model avoids overfitting to specific instances and instead captures shared patterns and relationships across tasks.
\textbf{For the downstream adaptation stage,} we design an adaptation algorithm called the Workload-adaptive Architectural Mask (WAM).
WAM provides a novel approach to knowledge transfer, emphasizing the inherent properties of the architecture rather than relying on workload similarity.

To summarize, we list our contributions as follows:
\begin{itemize}
    \item We conduct a comprehensive analysis of existing CPU cross-workload DSE frameworks, identifying their limitations in handling data ambiguity and enabling efficient knowledge transfer.

    \item We reframe the CPU cross-workload DSE task as a few-shot meta-learning problem, using the Model-Agnostic Meta-Learning (MAML) algorithm to enhance model adaptability with limited samples from new workloads.
    
    \item We introduce a novel adaptation algorithm named WAM, which captures inherent properties of the architecture, enabling efficient and generalized DSE framework across diverse workloads.
    
    \item Extensive experiments on the SPEC CPU 2017 benchmark suite show that our proposed MetaDSE framework significantly outperforms state-of-the-art methods, achieving 44.3\% reductions in prediction errors across various workloads.
\end{itemize}

%% file: Tex/2-Background.tex
\section{Background \& Related Work}

\subsection{Cross-workload DSE framework}
\subsubsection{Taxonomy}
Most cross-workload CPU DSE methods~\cite{TrEE, TrDSE, TrEnDSE, Microprocessor, ISCA19, PACT07, TC10} use transfer learning~\cite{transfer_learning, transfer_learning_survey} to reduce sample requirements for new workloads by leveraging source knowledge, typically following three strategies.
\textbf{Linear Fitting.} Work~\cite{TC10} uses linear regression to combine predictions from fixed source models, mapping them to the target workload label space using a few labeled samples. It merely builds a linear mapping between the source and target label spaces, relying heavily on the unrealistic assumption of identical distributions and linearity.
\textbf{Data Augmentation.} Work~\cite{ISCA19} employs the Gaussian mixture model~\cite{GMM} to model the original data and perform augmentation. By swapping the coefficients of high-probability and low-probability Gaussian components, the method enhances rare data distributions, aiming for more balanced data augmentation while preserving the overall distribution.
To some extent, this work stems from the premise of similarity in the distribution of workloads.
\textbf{Similarity Analysis.} Studies~\cite{Microprocessor, PACT07, TrEnDSE, TrDSE, TrEE} explore workload similarity for model adaptation using distinct strategies.
\cite{Microprocessor, PACT07} leverage workload signatures, representing workloads during pretraining and identifying the most similar signature for downstream adaptation. This approach transfers knowledge through signature-based representation.
In contrast, methods like TrEnDSE~\cite{TrEnDSE}, TrDSE~\cite{TrDSE}, and TrEE~\cite{TrEE} adopt a data-driven approach. TrEnDSE uses Wasserstein distance~\cite{Wasserstein} to measure workload similarity, while TrDSE clusters workloads using orthogonal array (OA) sampling and distributional features. TrEE further refines this with an OA-based foldover sampling strategy. These methods select similar samples from the pre-trained dataset to enhance downstream adaptation, directly integrating relevant upstream data without relying on explicit signature representation.

\subsubsection{Re-evaluating Similarity-based Approaches}

In this section, we examine the frameworks of incorporating data from source workload into downstream adaptation. One of the earliest works to apply this concept is TrDSE~\cite{TrDSE}, which influenced subsequent frameworks such as TrEE~\cite{TrEE} and TrEnDSE~\cite{TrEnDSE}. However, these methods always include a one-time training process using data from the target workload.
Actually, concerns over data leakage make it uncommon and impractical to access pre-trained data in many real-world scenarios.
These methods can be better described as data augmentation rather than transfer learning~\cite{transfer_learning_survey}. 

\subsection{Few-shot Problem \& Meta-learning}

The few-shot learning problem involves training models to generalize effectively using only a small number of labeled samples~\cite{few_shot_1, few_shot_2}, typically just a few dozen. This scenario is common in situations where labeling data is costly. To address this challenge, meta-learning has emerged as a powerful approach.
In meta-learning~\cite{meta_for_few_shot1, meta_for_few_shot1}, a model is trained across a diverse set of tasks and is optimized to generalize well over a distribution of tasks. Here, each task is associated with a dataset $\mathcal{D}$ containing features and labels, where the entire dataset acts as a single data sample. The goal of meta-learning can be formalized similarly to supervised learning, with the optimal model represented as $\theta^*=\arg\min_\theta\mathbb{E}_{\mathcal{D}\sim p(\mathcal{D})}[\mathcal{L}_\theta(\mathcal{D})]$.

\begin{figure}[!htbp] 
    \centering
    \vspace{-10pt}
    \includegraphics[width=0.47\textwidth]{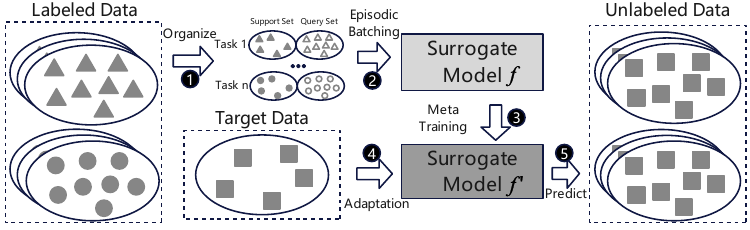}
    \vspace{-10pt}
    \caption{Typical workflow of Model-Agnostic Meta-Learning.}
    \label{fig:meta_workflow}
    \vspace{-10pt}
\end{figure}

As depicted in Fig.~\ref{fig:meta_workflow}, the meta-learning model is trained on a variety of curated tasks from different datasets, where the supervised signal from these tasks guides the model to perform well across a wide range of unseen tasks. One common meta-learning technique is MAML~\cite{MAML}, which aims to learn an optimal initialization of model parameters. This initialization enables the model to rapidly adapt to any new task with just a few gradient updates, providing efficient solutions for downstream few-shot problems.

%% file: Tex/3-Motivation.tex
\section{Motivation}

\subsection{Limitations in Prior Arts}
Current cross-workload DSE frameworks
have notable limitations in both the pre-training stage and the adaptation stage.

\subsubsection{Upstream Pre-training Issues}
In the pre-training stage, the surrogate model is prone to overfitting the pre-training data and faces data ambiguity. 
For instance, previous methods~\cite{TrEnDSE, TrDSE, TrEE} typically utilize machine learning based surrogate models, such as tree-based ones. These models tend to capture patterns specific to the source workloads and lead to overfitting with standard supervised training. 
However, due to the variability and ambiguity inherent in cross-workload CPU DSE scenarios, similar architectural configurations may yield vastly different performance metrics across workloads. This discrepancy leads the model to learn misleading or unreliable features during pre-training. Consequently, these features may not generalize well when the model is fine-tuned on new target workloads, resulting in poor predictive accuracy and limited adaptability.

\subsubsection{Downstream Adaptation Issues}
For the downstream adaptation stage, the primary challenge lies in the method of knowledge transfer. Many prior works assume that workloads are similar~\cite{Microprocessor, PACT07, TrEnDSE}, employing methods like Wasserstein distance~\cite{Wasserstein} to compare metric space distributions between source and target workloads, but this assumption does not always hold. 
In real-world scenarios, workloads exhibit significant variability in computational behaviors, data access patterns, and performance requirements, making this assumption unreliable.

\begin{figure}[!htbp] 
    \centering
    \vspace{-10pt}
    \includegraphics[width=0.47\textwidth]{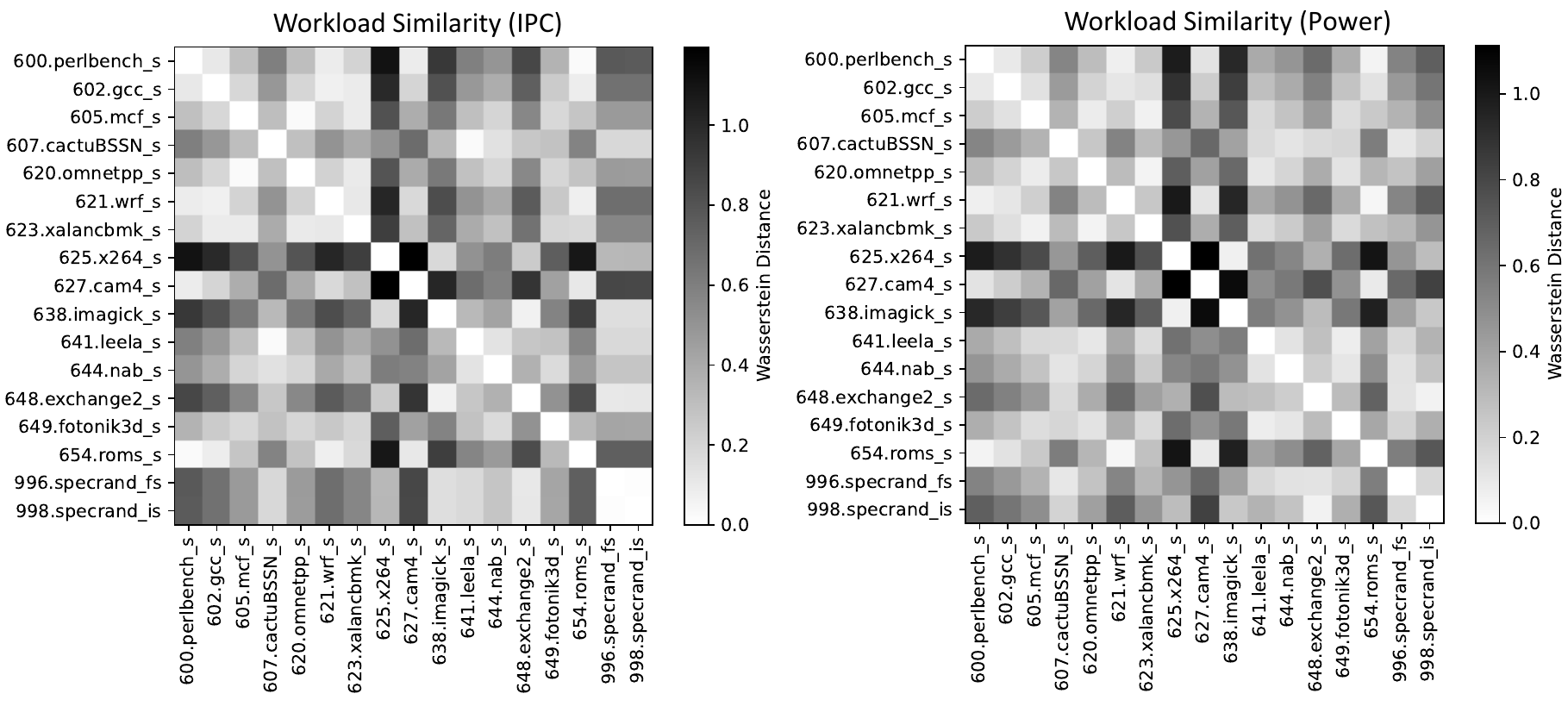}
    \vspace{-10pt}
	\caption{Wasserstein distances among SPEC CPU 2017 workloads. The darker the color, the less similar the two workloads.}
	\label{fig:workload_similarity}
    \vspace{-10pt}
\end{figure}

For instance, our experiments on the SPEC CPU 2017 benchmark suite (as shown in Fig.~\ref{fig:workload_similarity}) demonstrate that the similarities across workloads are inconsistent, with many showing substantial differences. This variability undermines the effectiveness of transfer learning methods like TrEnDSE when faced with diverse and complex workload characteristics. As a result, these methods may struggle to deliver accurate performance predictions for real-world DSE tasks, underscoring the need for a more adaptive, workload-independent approach to tackle the challenges of cross-workload CPU DSE.

\subsection{Rethinking Cross-workload CPU DSE}

To overcome the limitations of prior arts, we introduce two innovative solutions tailored to the distinct stages.

For the pre-training stage, we address the challenges of overfitting and data ambiguity by reframing the task as a few-shot meta-learning problem. This perspective shift allows the model to be trained on a variety of tasks using only a small number of labeled samples, thereby reducing the risk of overfitting and improving its ability to generalize across different workloads. By leveraging the meta-learning technique, the surrogate model can capture shared patterns across workloads, eliminating the impact of data ambiguity.

For the adaptation stage, we recognize the limitations of conventional methods that rely on workload similarity for knowledge transfer. Instead, we propose shifting the focus from workload similarity to the intrinsic properties of the architectural design. To harness these inherent characteristics, we introduce the WAM algorithm. WAM uncovers key relations between architectural parameters and leverages them during adaptation. This novel approach enhances the model’s robustness and adaptability, improving the overall effectiveness of cross-workload CPU DSE.

%% file: Tex/4-Methodology.tex
\section{Design of MetaDSE}

\subsection{Overview of MetaDSE}
\begin{figure}[!htbp] 
    \centering
    \vspace{-10pt}
    \includegraphics[width=0.47\textwidth]{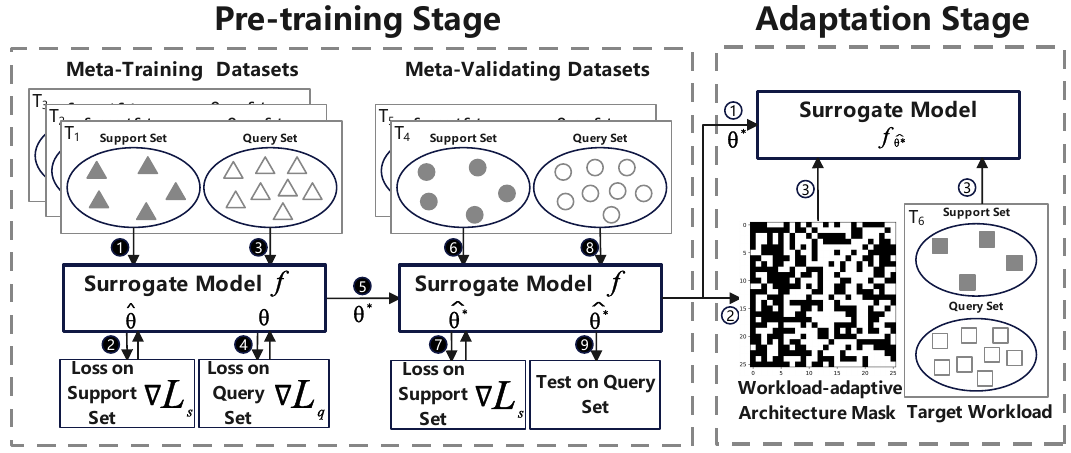}
    \vspace{-10pt}
    \caption{The overview workflow of MetaDSE.}
    \label{fig:workflow}
    \vspace{-10pt}
\end{figure}

In this section, we introduce MetaDSE, a framework designed to enhance the efficiency of cross-workload CPU DSE. 
The workflow of MetaDSE, illustrated in Fig.~\ref{fig:workflow}, is divided into two main stages: pre-training (\ding{182}–\ding{190}) and adaptation (\ding{172}–\ding{174}).
For the surrogate model $f$, we employ a transformer-based predictor, inspired by the strong performance of AttentionDSE~\cite{AttentionDSE} in CPU DSE tasks. 

In the pre-training stage, the meta-training process begins by creating a copy of the surrogate model (parameterized by $\hat{\theta}$). This model is first fine-tuned on the support set of each task through a few gradient descents (\ding{182}, \ding{183}). Subsequently, the fine-tuned model evaluates the loss on the corresponding query set (\ding{184}), and this computed loss contributes to the overall meta-loss (\ding{185}). The meta-loss is then used to update the parameters of the original surrogate model (parameterized by $\theta$). After each training epoch, a meta-validation step is conducted to evaluate the model’s performance and to identify the optimal parameters for downstream tasks.
In the adaptation stage, the pre-trained surrogate model (parameterized by $\hat{\theta^*}$) undergoes fine-tuning with the WAM algorithm (\ding{173}). WAM focuses on the inherent properties of the architecture, adjusting the model to emphasize relevant interactions between architectural parameters. Further technical details on the WAM adaptation process are elaborated in the subsequent sections.

\subsection{MAML-based Pre-training Stage}

The pre-training stage in MetaDSE typically aims to build a model that performs well across a wide range of tasks. Instead, our MAML-based approach focuses on finding an optimal model initialization, $\theta$, which can rapidly adapt to new downstream tasks using few data. This strategy allows the predictor to generalize effectively to diverse workloads without extensive retraining, tackling the core challenge of the few-shot problem in cross-workload CPU DSE.

\begin{algorithm}[!htbp]
    \SetAlgoLined
    \fontsize{8pt}{8pt}\selectfont
    \caption{\small \textbf{The Pre-Training Procedure of Predictors}}
    \label{alg:pretraining_predictor}
    \KwIn{$f_{\theta}$: the predictor with random parameters; $P(T)$: task distribution of one workload; $s, q$: sample amount of support set and query set in one task, correspondingly; $\alpha, \beta$: learning rate in inner and outer loop; }
    \KwOut{$f_{\theta^*}$: the meta-trained predictor}
    randomly initialize $\theta$ \\
    \While{not done}{
        Sample tasks from some workloads' distribution $T_i \sim P(T)$ \\
        \For{each task $t$ in $T_i$}{
            $\hat{\theta} = \theta$ \\
            Split out support set $S=(x_j,y_j)$ and query set $Q=(x_k, y_k)$ from $t$, $S, Q \leftarrow Spilt(t, s, q)$ \\
            \For{each inner step}{
                Evaluate $\nabla_{\hat{\theta}}\mathcal{L}(f_{\hat{\theta}})$ with respect to $S$ \\ 
                Compute adapted parameters with gradient: $\hat{\theta}=\hat{\theta} - \alpha \nabla_{\hat{\theta}}\mathcal{L}(f_{\hat{\theta}})$ 
            }
            Save $Q$ and $f_{\hat{\theta}}$ temporarily
        }
        Evaluate and average meta gradients of every task from $T_i$, $\nabla_{\theta}\mathcal{L}(f_{\hat{\theta}})$ with respect to each $Q$ in every $t$ \\
        Compute adapted meta parameters with meta gradient: $\theta=\theta-\beta \nabla_\theta \mathcal{L}_{T_i}(f_{\hat{\theta}})$ \\
    }
    $ \theta^* = \theta$ \\
    return final predictor $f_{\theta^*}$
\end{algorithm}

Algorithm~\ref{alg:pretraining_predictor} describes the MAML-based pre-training process for the surrogate model.
Before the training stage, each iteration begins by sampling tasks from the task distribution $P(T)$, representing different workloads. For each task, the model is trained using a support set $S$ and evaluated on a query set $Q$. 
The training in MAML can be divided into two phases: the inner loop and the outer loop.
The inner loop (from line 4 to line 12) focuses on task-specific adaptation, simulating the fine-tuning process during model deployment. Here, the model updates its parameters using the support set from a specific task. This phase aims to find temporary parameters $\hat{\theta}$ that adapt well to the current task. The update rule in this phase is: 
$\hat{\theta} \leftarrow \hat{\theta}-\alpha \nabla_{\hat{\theta}} \mathcal{L}(\hat{\theta})$.
where $\alpha$ is the inner loop learning rate. This gradient descent adjusts the parameters to minimize the task-specific loss $\mathcal{L}$, effectively mimicking how the model would adapt when exposed to new data.
The outer loop (from line 2 to line 15) aims to optimize the initial parameters 
$\theta$ by aggregating the feedback from multiple tasks. After the inner loop updates for each task, the model evaluates its performance on the query set to calculate the meta-loss. This meta-loss reflects how well the adapted model (with parameters 
$\hat{\theta}$) generalizes across different tasks. The meta-parameters $\theta$ are then updated based on the meta-gradient:
$\theta=\theta-\beta \nabla_\theta \mathcal{L}_{T_i}(f_{\hat{\theta}})$
where $\beta$ is the meta-learning rate. This outer-loop update ensures that the initial parameters $\theta$ are optimized to quickly adapt to new tasks with minimal data.

This two-level optimization process enables MetaDSE to efficiently adapt to new workloads. It ensures the pre-trained model generalizes well to diverse, unseen tasks, minimizing the need for extensive fine-tuning on large datasets.

\subsection{Workload-adaptive Architectural Mask Adaptation Stage}

In cross-workload CPU DSE tasks, workload dissimilarity poses a significant challenge for knowledge transfer. To tackle this, we propose WAM adaptation, which leverages inherent architectural properties to enhance the pre-trained model’s generalization to unseen workloads.

A critical issue is that not all architectural parameters are fully correlated. Previous efforts~\cite{TrEnDSE, TrDSE, TrEE, PACT07, TC10} have largely overlooked this, leading to noise from irrelevant parameter interactions and reduced prediction accuracy. To address this, we introduce an algorithm that generates a mask for the surrogate model, utilizing intermediate attention weights from the transformer-based predictor during pre-training. This mask filters out irrelevant parameter interactions, enabling the model to focus on key factors and improving prediction accuracy.

\begin{figure}[!htbp] 
    \centering
    \includegraphics[width=0.4\textwidth]{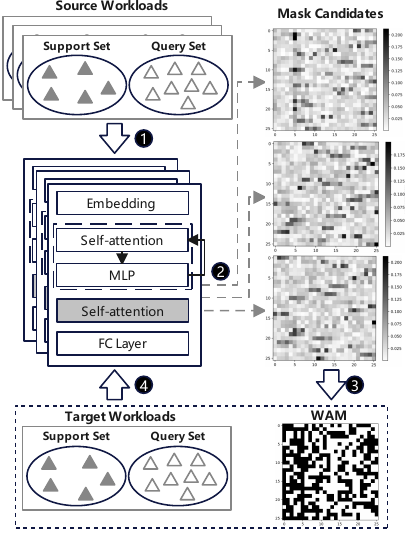}
    \vspace{-10pt}
    \caption{The example of workload-adaptive architectural mask generation.}
    \label{fig:WAM}
\end{figure}

As Fig.~\ref{fig:WAM} shows, we utilize attention weights obtained during the pre-training process to analyze the relationships between architectural parameters (\ding{182}, \ding{183}). 
We extract all the attention weights from the last self-attention layer, referring to them as mask candidates.
By identifying parameter interactions that frequently occur across diverse workloads, we construct a WAM (\ding{184}) that highlights high-frequency correlations, effectively filtering out less relevant interactions.

\begin{algorithm}[!t]
    \SetAlgoLined
    \fontsize{8pt}{8pt}\selectfont
    \caption{\small \textbf{The Adaptation Procedure of Predictors}}
    \label{alg:adaptation_predictor}
    \KwIn{$f_{\theta^*}$: the meta-trained predictor; $P(T)$: task distribution of test workload; $s, q$: sample amount of support set and query set in one task, correspondingly; $\gamma$: learning rate during adaptation; $M$: workload-adaptive architectural mask.}
    \KwOut{$f_{\hat{\theta^*}}$: the adapted predictor}
    Equip $f_{\theta^*}$ with $M$ in its self-attention operator \\
    Set $M$ to learnable state: $M.required\_grad = True$ \\
    Sample tasks from test workload's distribution $T_i \sim P(T)$ \\
    \For{each task $t$ in $T_i$}{
        $\hat{\theta^*} = \theta^*$ \\
        Split out support set $S=(x_j,y_j)$ and query set $Q=(x_k, y_k)$ from $t$, $S, Q \leftarrow Spilt(t, s, q)$ \\
        \For{each step}{
            Evaluate $\nabla_{\hat{\theta^*}}\mathcal{L}(f_{\hat{\theta^*}}, M)$ with respect to $S$ \\ 
            Compute adapted parameters with gradient: $\hat{\theta^*}=\hat{\theta^*} - \gamma \nabla_{\hat{\theta^*}}\mathcal{L}(f_{\hat{\theta^*}})$ 
        }
        return adapted predictor $f_{\hat{\theta^*}}$
    }
\end{algorithm}

After the mask is generated, we leverage the mask and the target workloads to the adaptation stage (\ding{185}). 
The procedure is illustrated in Algorithm~\ref{alg:adaptation_predictor}.
The process also begins by sampling tasks from the test workload's distribution $P(T)$. For each sampled task $t$, we initialize the adapted model parameters as $\hat{\theta^*} = \theta^*$ and split the task data into a support set $S$ and a query set $Q$. During the adaptation process, we iteratively update the model parameters $\hat{\theta^*}$ using gradients computed based on the loss $\mathcal{L}$ evaluated on the support set $S$. The update rule $\hat{\theta^*}=\hat{\theta^*} - \gamma \nabla_{\hat{\theta^*}}\mathcal{L}(f_{\hat{\theta^*}})$ ensures that the model learns with mask $M$ to minimize the error on the specific task at hand.
Finally, the adapted predictor $f_{\hat{\theta^*}}$ is returned for each task, incorporating the updated parameters that better capture the features of target workloads. 

WAM adaptation allows the surrogate model to focus on critical parameter interactions, reducing noise from irrelevant features and enhancing prediction accuracy. Moreover, it introduces a novel approach to knowledge transfer, prioritizing architectural properties over workload similarity.

% The WAM adaptation enables the surrogate model to emphasize critical parameter interactions when adapting to new workloads, effectively reducing noise from irrelevant features and improving prediction accuracy. Additionally, WAM offers a novel perspective for knowledge transfer, focusing not on workload similarity but on the inherent properties of the architecture.

%% file: Tex/5-Experiment_Setup.tex
\section{Experimental Setup}

% In this section, we introduce the experimental setup used to assess the performance of MetaDSE.

\textbf{Datasets Generation.}
We utilize GEM5~\cite{gem5} as the timing-accurate simulator and McPAT~\cite{mcpat} as the power modeling tool. We extend GEM5 using Python to generate different CPU cores with varying configurations.
To bolster the credibility of our experiments, we employ SPEC CPU 2017~\cite{SPEC2017} as the evaluation benchmark. 
We utilize Simpoints~\cite{simpoints} for each workload evaluation. Each workload is divided into at most 30 clusters, with each cluster containing ten million instructions.

For dataset splitting, we iteratively and randomly designated seven datasets for training, five for validation, and five for testing. This approach ensured balanced and robust training across various dataset combinations.

\textbf{Baselines.}
We evaluate our proposed MetaDSE framework against several baselines from different perspectives. The first baseline is TrEnDSE~\cite{TrEnDSE}, a state-of-the-art cross-workload CPU DSE framework. 
We then introduce TrEnDSE-Transformer, where the ensemble model in TrEnDSE is replaced with a Transformer-based predictor.
Lastly, MetaDSE-w/o WAM is a variant of MetaDSE that excludes the WAM adaptation.

\textbf{The Explored Design Space.}
The design space of the out-of-order processor is listed in Table~\ref{tb:design_space}. The values in the third column are formatted as ``start number: end number: stride". A two-level cache hierarchy with an 8192MB DRAM main memory configuration is employed.

\begin{table}[!htbp]
\centering
\fontsize{8pt}{10pt}\selectfont
\caption{Microarchitecture design space specification.}
\vspace{-5pt}
\label{tb:design_space}
\begin{tabular}{|c|cc|}
\hline
\textbf{Parameters} & \multicolumn{1}{c|}{\textbf{Description}}                                                                                    & \textbf{\begin{tabular}[c]{@{}c@{}}Candidate\\ Value\end{tabular}} \\ \hline
Core Frequency      & \multicolumn{1}{c|}{\begin{tabular}[c]{@{}c@{}}the frequency of\\ CPU core in GHz\end{tabular}}                              & \begin{tabular}[c]{@{}c@{}}1/1.5/2/\\ 2.5/3\end{tabular}           \\ \hline
Pipeline Width      & \multicolumn{1}{c|}{\begin{tabular}[c]{@{}c@{}}fetch/decode/rename/\\ dispatch/issue/writeback/\\ commit width\end{tabular}} & 1:12:1                                                             \\ \hline
Fetch Buffer        & \multicolumn{1}{c|}{\begin{tabular}[c]{@{}c@{}}fetch buffer size\\ in bytes\end{tabular}}                                    & 16/32/64                                                           \\ \hline
Fetch Queue         & \multicolumn{1}{c|}{\begin{tabular}[c]{@{}c@{}}fetch queue size\\ in $\mu$-ops\end{tabular}}                                 & 8:48:4                                                             \\ \hline
Branch Predictor      & \multicolumn{1}{c|}{predictor type}                                                                                          & \begin{tabular}[c]{@{}c@{}}BiModeBP/\\ TournamentBP\end{tabular}   \\ \hline
% Choice Predictor    & \multicolumn{1}{c|}{choice predictor size}                                                                                   & 2048/4096/8192                                                     \\ \hline
% Global Predictor    & \multicolumn{1}{c|}{global predictor size}                                                                                   & 2048/4096/8192                                                     \\ \hline
RAS Size            & \multicolumn{1}{c|}{return address stack size}                                                                               & 16:40:2                                                            \\ \hline
BTB Size            & \multicolumn{1}{c|}{branch target buffer size}                                                                               & 1024/2048/4096                                                     \\ \hline
ROB Size            & \multicolumn{1}{c|}{reorder buffer entries}                                                                                  & 32:256:16                                                          \\ \hline
Int/Fp RF Number       & \multicolumn{1}{c|}{\begin{tabular}[c]{@{}c@{}}number of physical\\ integer/floating-point registers\end{tabular}}                            & 64:256:8                                                           \\ \hline
Inst Queue          & \multicolumn{1}{c|}{\begin{tabular}[c]{@{}c@{}}number of instruction\\ queue entries\end{tabular}}                           & 16:80:8                                                            \\ \hline
Load/Store Queue          & \multicolumn{1}{c|}{\begin{tabular}[c]{@{}c@{}}number of\\ load/store queue entries\end{tabular}}                                  & 20:48:4                                                            \\ \hline
IntALU              & \multicolumn{1}{c|}{number of integer ALUs}                                                                                  & 3:8:1                                                              \\ \hline
IntMultDiv          & \multicolumn{1}{c|}{\begin{tabular}[c]{@{}c@{}}number of integer\\ multipliers and dividers\end{tabular}}                    & 1:4:1                                                              \\ \hline
FpALU               & \multicolumn{1}{c|}{\begin{tabular}[c]{@{}c@{}}number of \\ floating-point ALUs\end{tabular}}                                & 1:4:1                                                              \\ \hline
FpMultDiv           & \multicolumn{1}{c|}{\begin{tabular}[c]{@{}c@{}}number of floating-point\\ multipliers and dividers\end{tabular}}             & 1:4:1                                                              \\ \hline
Cacheline           & \multicolumn{1}{c|}{cacheline size}                                                                                          & 32/64                                                              \\ \hline
L1 Cache Size      & \multicolumn{1}{c|}{size of ICache in KB}                                                                                    & 16/32/64                                                           \\ \hline
L1 Cache Assoc.    & \multicolumn{1}{c|}{\begin{tabular}[c]{@{}c@{}}associative sets of \\ ICache\end{tabular}}                                   & 2/4                                                                \\ \hline
L2 Cache Size       & \multicolumn{1}{c|}{size of L2 Cache in KB}                                                                                  & 128/256                                                            \\ \hline
L2 Cache Assoc.     & \multicolumn{1}{c|}{\begin{tabular}[c]{@{}c@{}}associative sets of \\ L2 Cache\end{tabular}}                                 & 2/4                                                                \\ \hline
\end{tabular}
\vspace{-10pt}
\end{table}

\textbf{Evaluation Metrics.}
The accuracy of the model is primarily measured using the Root Mean Squared Error (RMSE), which is calculated as:
\begin{equation}
    \text { RMSE }=\sqrt{\frac{1}{N} \sum_{i=1}^{N}\left(y_{i}-\hat{y}_{i}\right)^{2}}
\end{equation}
where $y_i$ is the actual value, $\hat{y}_{i}$ is the predicted value, and $N$ is the number of data points.

Additionally, other metrics such as Mean Absolute Percentage Error (MAPE) and Explained Variance (EV) are used for a more comprehensive evaluation:

\begin{equation}
    \text { MAPE }=\frac{1}{n} \sum_{i=1}^{n}\left(\frac{\left|y_{i}-y_{i}^{\text {real }}\right|}{y_{i}^{\text {real }}}\right) \times 100.
\end{equation}

\begin{equation}
    \mathrm{EV}=1-\frac{\sum_{i=1}^{N}\left(y_{i}-\hat{y}_{i}\right)^{2}}{\sum_{i=1}^{N}\left(y_{i}-\bar{y}\right)^{2}}
\end{equation}
where $\bar{y}$ is the mean of the actual values. Lower RMSE and MAPE values, and higher EV values, indicate better model performance.

%% file: Tex/6-Experiment.tex
\section{Experiment Results}

\subsection{Overall Analysis}
\begin{figure*}[!htbp] 
	\centering
	\includegraphics[width=0.97\textwidth]{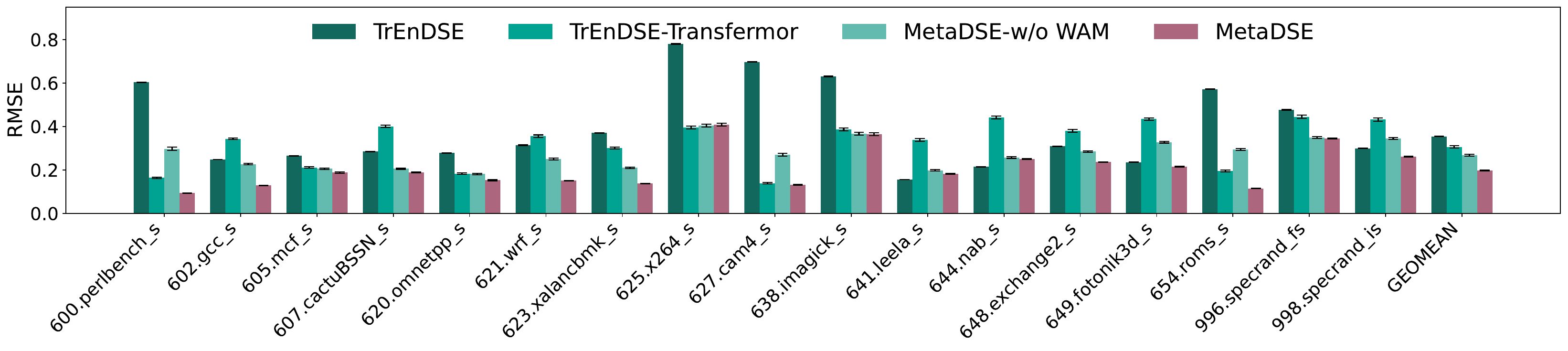}
	\vspace{-10pt}
	\caption{Comparison of IPC RMSE per workload on SPEC CPU 2017 with the SOTA cross-workload CPU DSE framework (lower is better for RMSE).}
	\label{fig:RMSE}
    \vspace{-10pt}
\end{figure*}

\begin{table*}[!htbp]
\centering
\fontsize{9pt}{10pt}\selectfont
\caption{The overall results, averaged across five datasets (600.perlbench\_s, 605.mcf\_s, 620.omnetpp\_s, 623.xalancbmk\_s, and 627.cam4\_s), are evaluated using RMSE, MAPE, and Explained Variance.}
\label{tb:multi_metrics}
\begin{tabular}{@{}ccccccc@{}}
\toprule
\multirow{2}{*}{\textbf{Models / Metrics}} & \multicolumn{2}{c}{\textbf{RMSE $\downarrow$}} & \multicolumn{2}{c}{\textbf{MAPE $\downarrow$}} & \multicolumn{2}{c}{\textbf{EV $\uparrow$}} \\ \cmidrule(l){2-7} 
 & \textbf{IPC} & \textbf{Power} & \textbf{IPC} & \textbf{Power} & \textbf{IPC} & \textbf{Power} \\ \midrule
\multicolumn{1}{c|}{\textbf{RF}} & \acc{0.4389}{0.0150} & \acc{0.5344}{0.0198} & \acc{1.1624}{0.0827} & \acc{0.3356}{0.0178} & \acc{-0.7997}{0.1231} & \acc{0.4470}{0.0509} \\
\multicolumn{1}{c|}{\textbf{GBRT}} & \acc{0.3637}{0.0125} & \acc{0.4539}{0.0171} & \acc{0.9486}{0.0678} & \acc{0.2667}{0.0144} & \acc{-0.5152}{0.1137} & \acc{0.4634}{0.0529} \\
\multicolumn{1}{c|}{\textbf{TrEnDSE}} & \acc{0.3270}{0.0109} & \acc{0.3990}{0.0146} & \acc{0.8386}{0.0595} & \acc{0.2348}{0.0123} & \acc{-0.5142}{0.1079} & \textbf{\acc{0.5711}{0.0410}} \\
\multicolumn{1}{c|}{\textbf{MetaDSE}} & \textbf{\acc{0.2204}{0.0072}} & \textbf{\acc{0.3969}{0.0095}} & \textbf{\acc{0.5909}{0.0460}} & \textbf{\acc{0.2330}{0.0045}} & \textbf{\acc{-0.0471}{0.0056}} & \acc{0.3189}{0.0065} \\ \bottomrule
\end{tabular}
\vspace{-10pt}
\end{table*}

The implementation of MetaDSE has two stages: upstream pre-training and downstream WAM adaptation. 
For pre-training, we run 15 epochs with 200 tasks per workload, each task having five support samples and forty-five query samples. Inner and outer loop learning rates are $1 \times 10^{-5}$ and $1 \times 10^{-4}$, respectively. 
In each iteration, every model copy is optimized via five inner-loop steps on the support set using SGD, and the accumulated meta-loss is used to update the original model with the Adam optimizer.
In the downstream stage, we fine-tune the pre-trained model using ten gradient steps with a learning rate of $1 \times 10^{-5}$ and cosine annealing. We evaluate 1000 tasks per workload for mean and confidence intervals.

Fig.~\ref{fig:RMSE} illustrates the prediction accuracy comparison between MetaDSE and various baseline models on the SPEC CPU 2017 dataset. The last column presents the geometric mean (GEOMEAN) of prediction errors across all tasks.
MetaDSE achieves a significant reduction in prediction error, lowering it by 44.3\% compared to the state-of-the-art cross-workload CPU DSE framework, TrEnDSE. 
When comparing TrEnDSE with TrEnDSE-Transformer, we observe that simply replacing the ensemble learning model with a more powerful Transformer model provides some performance improvement. However, this improvement is relatively modest and inconsistent across different tasks. 
In contrast, within the meta-learning-based DSE framework, the inclusion of our proposed WAM adaptation technique leads to a substantial performance boost, reducing the average prediction error by 27\%, and highlighting the effectiveness of adapting to workload-specific characteristics.

We further compare MetaDSE with several commonly used models in transfer learning for predicting IPC and power, including Random Forest (RF) and Gradient Boosting Regression Trees (GBRT), shown in Table~\ref{tb:multi_metrics}. 
Our results reveal that MetaDSE consistently outperforms TrEnDSE and other baselines across all evaluated workloads, demonstrating lower RMSE and MAPE values while achieving higher EV scores.
These results further highlight the superiority of MetaDSE over traditional transfer learning approaches, emphasizing its enhanced ability to generalize and adapt effectively across different workloads in cross-workload CPU DSE tasks.

\subsection{Exploring Data Impact}
We further investigate the impact of pre-training and adaptation data on MetaDSE performance.

\subsubsection{Sensitivity of Pre-training}
\begin{figure}[!htbp] 
	\centering
	\includegraphics[width=0.47\textwidth]{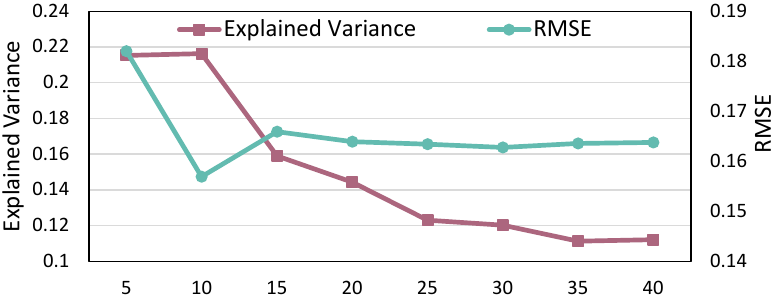}
	\vspace{-10pt}
	\caption{The variation in explained variance and RMSE with respect to changes in the source sample size.}
	\label{fig:sensitive_of_source}
\end{figure}

For the upstream pre-training stage, we fix the downstream adaptation support set size to ten and change the support set size of the pre-training stage from five to forty. 
Fig.~\ref{fig:sensitive_of_source} illustrates that when the upstream and downstream data are aligned, the model achieves optimal transfer performance, with the highest EV and lowest RMSE. 
This improvement is attributed to the matched knowledge pattern and the higher learning efficiency when the data distribution is consistent across pre-training and adaptation phases, leading to better generalization and model adaptation.

\subsubsection{Sensitivity of Adaptation}
We further investigate the impact of downstream adaptation data and compare MetaDSE to commonly used machine learning models, including RF, GBRT, and TrEnDSE. In this experiment, we fix the support set size of the upstream pre-training stage to ten while varying the adaptation support set size from five to forty. The results presented in Table~\ref{tb:sensitive_of_adaptation} show that when the pretraining data size is fixed, increasing the adaptation data size results in modest performance improvements for MetaDSE. Notably, MetaDSE achieves high performance even with a smaller amount of adaptation data.

\begin{table}[]
\centering
\fontsize{9pt}{10pt}\selectfont
\caption{RMSE on IPC prediction as the downstream adaptation support size (K) changes.}
\label{tb:sensitive_of_adaptation}
\begin{tabular}{c|cclccc}
\toprule
\textbf{Models / K} & \textbf{5} & \multicolumn{2}{c}{\textbf{10}} & \textbf{20} & \textbf{30} & \textbf{40} \\ \midrule
\textbf{RF} & 0.4409 & \multicolumn{2}{c}{0.4397} & 0.4390 & 0.4386 & 0.4380 \\
\textbf{GBRT} & 0.2577 & \multicolumn{2}{c}{0.2390} & 0.2356 & 0.2321 & 0.2299 \\
\textbf{Baseline} & 0.2616 & \multicolumn{2}{c}{0.2397} & 0.2229 & 0.2147 & 0.2076 \\
\textbf{MetaDSE} & 0.1580 & \multicolumn{2}{c}{0.1562} & 0.1485 & 0.1471 & 0.1466 \\ \bottomrule
\end{tabular}
\vspace{-10pt}
\end{table}

%% file: Tex/7-Conclusion.tex
\section{Conclusion}
In this work, we introduce meta-learning to cross-workload CPU DSE tasks and propose MetaDSE. MetaDSE utilizes the MAML and proposes the WAM algorithm to uncover the inherent architectural properties, offering a novel approach to prior knowledge transfer. Experiments on SPEC CPU 2017 benchmarks demonstrate that MetaDSE significantly reduces prediction errors compared to state-of-the-art methods, highlighting its effectiveness in improving model adaptability.